\def \cm{~\rm{cm}}
\def \s{~\rm{s}}
\def \km{~\rm{km}}

\def \AU{~\rm{AU}}
\def \erg{~\rm{erg}}

\def \yr{~\rm{yr}}
\def \pc{~\rm{pc}}

\documentclass[12pt,preprint]{aastex}
\usepackage{natbib}

\shorttitle{FORMATION OF WIDE JETS}
\shortauthors{Soker}
\slugcomment{Draft version of \today}

\begin{document}

\title{THE FORMATION OF SLOW-MASSIVE-WIDE JETS}

\author{Noam Soker\altaffilmark{1}}

\altaffiltext{1}{Dept. of Physics, Technion, Haifa 32000, Israel;
soker@physics.technion.ac.il.}

\begin{abstract}
I propose a model for the formation of slow-massive-wide (SMW) jets by accretion
disks around compact objects.
This study is motivated by claims for the existence of SMW jets in some astrophysical objects
such as in planetary nebulae (PNs) and in some active galactic nuclei in galaxies and
in cooling flow clusters.
In this model the energy still comes from accretion onto a compact object.
The accretion disk launches two opposite jets with velocity of the order of the escape velocity
from the accreting object and with mass outflow rate of $\sim 1-20 \%$ of the accretion
rate as in most popular models for jet launching; in the present model these are termed
fast-first-stage (FFS) jets.
However, the FFS jets encounter surrounding gas that originates in the mass accretion
process, and are terminated by strong shocks close to their origin.
Two hot bubbles are formed. These bubbles accelerate the surrounding gas to form two SMW jets
that are more massive and slower than the FFS jets.
There are two conditions for this mechanism to work.
Firstly, the surrounding gas should be massive enough to block the free expansion of the FFS jets.
Most efficiently this condition is achieved when the surrounding gas is replenished.
Secondly, the radiative energy losses must be small.
\end{abstract}

\section{INTRODUCTION}
\label{sec:intro}

Jets are usually associated with narrowly collimated outflows, e.g.,
half opening angle of $\alpha \la 10^\circ$. For example, in the majority of
numerical simulations the jets are taken to have cylindrical outflow, namely,
$\alpha=0^\circ$. In recent year there are more indications and suggestions
that outflows with wide opening angle occur in some systems, from progenitors
of planetary nebulae (PNs) to active galactic nuclei (AGN).
Most prominent are observational indications for massive wide jets in AGNs
(Behar et al. 2003; Crenshaw \& Kraemer 2007; de Kool et al. 2001;
Kaspi \& Behar 2006).
Wide jets were introduced into numerical simulations by Sternberg et al. (2007) to
explain the inflation of fat (more or less sphreical) bubbles attached to the center of
clusters of galaxies.
Massive-slow, but not wide, jets were considered and simulated by several groups in
recent years (e.g., Begelman \& Celotti 2004; Binney 2004;
Omma et al. 2004; Heinz et al. 2006; {{{ Vernaleo \& Reynolds 2007). }}}
Massive jets, for example, might be more efficient in maintaining a feedback heating
in cooling flow clusters of galaxies (Soker \& Pizzolato 2005).

Wide jets were considered (Soker 2004a) and simulated (Akashi 2007; Akashi \& Soker 2008) for
inflating fat bubbles in PNs in a process similar to that in clusters of galaxies.
In some PNs the motion of the nebula is ballistic, i.e., the outflow velocity of
different nebular segments is proportional to their distance from the center
(e.g., Sanchez Contreras et al. 2006).
This type of flow indicates that most of the nebular mass was ejected during a
relatively short time. Namely, a massive outflow, probably in the form of jets
launched from the accretion disk around a companion.
In some PNs the jets are narrow, but in others might be wide.

The question addressed in this paper is what are the conditions under which an outflow
launched from the inner zone of an accretion disk can turn into a slow-massive-wide (SMW) jet.
The fast outflow from the inner zone of the accretion disk is termed here
the fast-first-stage (FFS) jets, as there are two oppositely launched jets.
In this paper I will consider wide jets formed by the FFS jets as they interact
with a surrounding gas, and will term them slow-massive-wide (SMW) jets,
where by `slow' it is understood that the jet terminal sped is much lower than the escape
speed from the accreting object and from the velocity of the FFS jets, and by `massive' it
is understood that the outflowing mass in the SMW jets is comparable to, or larger than, the
accretion rate into the central accreting object.
This type of outflow, although very different quantitatively,
was simulated recently in a different setting by Sternberg et al. (2007) and
Sutherland \& Bicknell (2007).
Sutherland \& Bicknell (2007) simulated the interaction of jets with the ISM
in a disk inside elliptical galaxies.
In their simulations the initial jets are cylindrical. The initial interaction between
the jets and ISM occurs very close to the center, at about few jet's diameter.
Practically, therefore, the jets are wide jets. In addition, the jet velocity
in their Model A is $\sim 13,500 \km \s^{-1}$ and the mass loss rate into two jets is
$\sim 1 M_\odot \yr^{-1}$.
Therefore, without stating it explicitly, they practically simulated slow-massive-wide
jets. For that, it is not surprising that at early times they inflate
hot bubbles (Soker 2004a) as in the gasdynamical simulations of wide jets
conducted by Sternberg et al. (2007).
Later in the simulations of Sutherland \& Bicknell (2007) the initial jets penetrate the ISM,
whereafter they are collimated by the ISM and form narrow jets.
The simulations of Sternberg et al. (2007) and Sutherland \& Bicknell (2007) show that in
principle jets can accelerate surrounding gas into a wide slower outflow.
In the present paper the initial setting is very different,
but the physical mechanism is somewhat similar.

In section 2 I consider the physical conditions near the accretion disk, while
in section 3 I consider the constraints on the FFS jets.
A short summary is in section 4.

\section{THE CONSTRAINTS ON THE ACCRETION PROCESS}
\label{sec:conditions}

\subsection{The basic assumptions and geometry}
I consider the formation of SMW (slow-massive-wide) jets when the mass transfer rate to
the accretion disk is large, and in addition the central accreting object might have a
relative speed to the source supplying the mass.
For example, in a binary system the accreting companion
orbit the mass donor star, while the mass transfer rate is so high that the
mass transfer occurs not only via a roche lobe overflow (RLOF) through the first
Lagrangian point (L1), but also by a wind concentrated around L1.
The high mass transfer rate and the relative motion of the accreting object relative
to the mass source will lead to the presence of gas above and below the accreting object,
namely along the symmetry axis of the accretion disk.
In particular, if the mass inflow rate into the disk increases on a time scale shorter
than the viscous time scale, the disk will not have time to rearrange itself, and the
large departure from a steady state might lead to mass expelled to above and below the
accretion disk.

The large mass transfer rate implies higher density in the disk, hence higher
optical depth; the material might be optically thick above and below the disk,
and not only in the disk.
The energy released in the accretion disk by dissipation can uplift mass below and
above the disk, but in general cannot eject a massive jet by itself,
as will be shown below.

It is in this outset that the two jets launched from the central zone of the disk,
what are termed here the fast-first-stage-jets (FFS jets)
might eject a large mass out, and form a wide massive jet, which by energy conservations will
be much slower than the FFS jets, i.e., much below the escape speed from the accreting object.
The initial geometry and flow structure are drawn schematically in Figure 1.
\begin{figure}
\vskip -1.5 cm
\resizebox{0.55\textwidth}{!}{\includegraphics{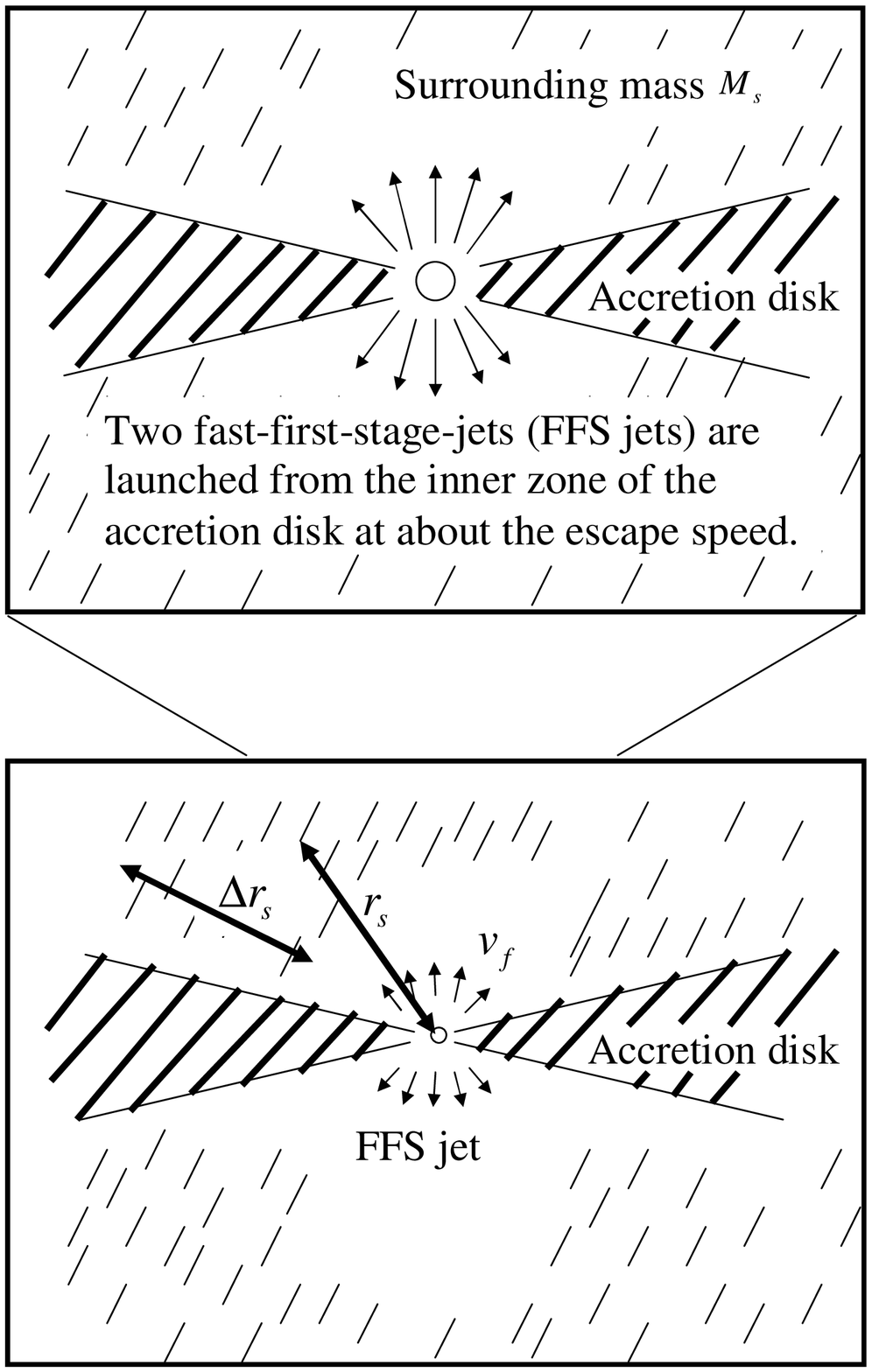}}
\vskip -0.0 cm
\caption{Schematic drawing of the initial geometry and flow structure.
High mass transfer rate and relative motion of the accreting body and
the mass source lead to the presence of mass $M_s$ above and below the
entire mid plane, including the symmetry axis of the accretion disk.
A fast outflow, the fast-first-stage-jets (FFS jets), is launched from the inner zone
of the accretion disk.
When not much mass resides along the symmetry axis and it is not replenished,
the FFS jets from the inner zone penetrate the mass $M_s$ and are collimated by
the surrounding mass and form two opposite narrow jets.
In the present paper the conditions are assumed to be such that the FFS jets cannot
penetrate the material along the symmetry axis for a long enough time, and instead
the FFS jets accelerate the mass $M_s$ and form SMW jets.
}
\label{initial}
\end{figure}

Let us assume that a fraction $k<1$ of the energy released locally by the disk
is channelled to uplift or eject mass from the disk.
The energy released locally by the disk in an annulus $dr$ and per unit time is
$dE/dt=(d \dot E/dr)dr=(G M /2r) (dr/r) \dot M_{\rm acc}$, where
$\dot M_{\rm acc}$ is the accretion rate through radius $r$,
and $M$ is the mass of the accreting object.
Let $d \dot M_w$ be the mass ejection rate from the annulus $dr$,
and with a terminal speed $v_w=\beta v_K$, where $v_K=(GM/r)^{1/2}$ is
the Keplerian velocity at $r$.
Energy conservation reads
\begin{equation}
\frac{1}{2} k \left(\frac{GM}{r} \right) \frac {dr}{r} \dot M_{\rm acc}
=\left( \frac{1}{2} \frac{GM}{r} + \frac{1}{2} v_w^2 \right) d \dot M_w.
\label{enn1}
\end{equation}
Using the mass conservation $\dot M_{\rm acc}=-d \dot M_w$, equation (\ref{enn1})
can be solved to yield
\begin{equation}
M_{\rm acc}(r) =M_{\rm acc0} \left( \frac{r}{r_0} \right)^{-k/(1+\beta^2)},
\label{eq:enn2}
\end{equation}
where $M_{\rm acc0}$ is the mass accretion rate through radius $r_0 > r$.
If the material above the disk is optically thin, then $k<<1$ and most of the mass
injected into the disk at $r_0$ reaches small radii $r \ll r_0$.
The small amount that the disk does eject forms a disk wind. This wind is required to collimate
narrow jets, as a confining external pressure is always required at the edge of
accelerated jets (Komissarov et al. 2007).

If, on the other hand, the accretion rate is very high and mass flows above and below the disk,
and we require the disk only to uplift material, namely, $\beta \ll 1$, then
we have $k/(1+\beta^2) \sim 1$, and the disk can uplift a large fraction of the injected mass.
For example, if $k/(1+\beta^2)=1/2$, then the mass accreted at $r=0.01 r_0$ is only $\sim 10 \%$ of
the mass injected into the disk. The rest of the mass resides above and below the disk.
This mass, by my assumption, has a very low terminal speed and will not be ejected from
the system at a high speed and will not form a jet, but rather a disk wind.
In the present paper I investigate the case where mass exist also along the symmetry axis of
the accretion disk; it can come from the outer region of the accretion disk and/or directly
from the mass transfer process itself.
In that case the FFS jets launched from the inner zone of the disk
cannot penetrate the dense material.
Instead the FFS jets form two hot bubbles that accelerate the disk wind and the material that
reaches directly the regions above and below the disks (see below), and a slower
outflow, the SMW jets, is formed.
This is schematically drawn in Figure 2.
\begin{figure}
\vskip -3.5 cm
\resizebox{0.77\textwidth}{!}{\includegraphics{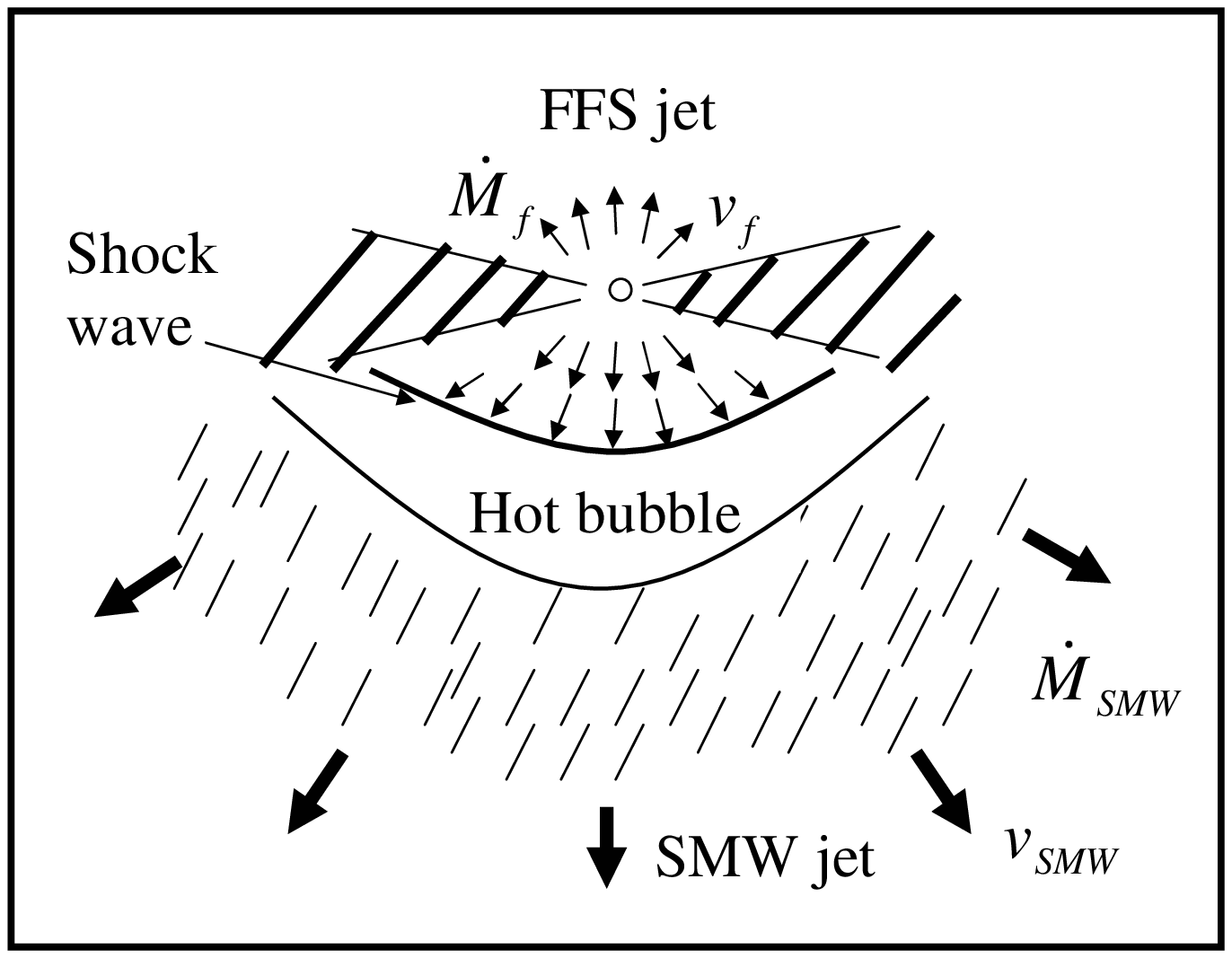}}
\vskip -0.0 cm
\caption{Schematic drawing of the formation of the slow-massive-wide (SMW) jets,
drawn only in the lower half of the meridional plane.
Two fast-first-stage (FFS) jets are launched from the inner zone of the accretion disk
with a velocity $v_f$ about equal to the escape velocity from the accreting object,
and a mass loss rate $\dot M_f$ of $\sim 1-20 \%$ of the accretion rate.
If the conditions for not penetrating the surrounding gas and for a long radiative cooling
time are met, then a hot bubble that accelerates the surrounding gas is formed.
The surrounding gas mass is large such that the final SMW jets have mass loss rate of
$\dot M_{SMW} \gg \dot M_f$, and from energy conservation their speed $v_{\rm SMW} \ll v_f$.
}
\label{SMWjets}
\end{figure}

\subsection{The material along the symmetry axis}

The inner zone of the accretion disk launches fast outflow into the two sides of the
disk, the FFS jets. If the FFS jets penetrate the gas around the accretion disk they
will be collimated by that gas, and two narrow collimated fast jets will be formed,
similar to the flow structure in the simulations of Sutherland \& Bicknell (2007).
By fast it is understood that the jet's speed is of the order of the escape speed
from the inner zone of the disk.
If, on the other hand, the FFS jets cannot penetrate the surrounding gas they will
accelerate the surrounding gas and form SMW (slow-massive-wide) jets.
In this section the conditions for the FFS jets not to penetrate the surrounding gas
are derived.

Let the FFS jets from the inner disk zone have a mass outflow rate to both sides
of $\dot M_f$, a velocity $v_f$, and let the two FFS jets cover a solid angle of
$4 \pi \delta$ (on both sides of the disk together).
The density of the outflow at radius $r$ is
\begin{equation}
\rho_f =  \frac {\dot M_f}{4 \pi \delta r^2 v_f}.
\label{eq:rhof}
\end{equation}
Let the FFS jets encounter the surrounding gas residing at distance
$r_s$ in a shell of width $\Delta r_s \simeq r_s$, and density $\rho_s$.
The head of each jet proceeds at a speed $v_h$ given by the balance
of pressures on its two sides.
Assuming supersonic motion this equality reads
$\rho_s v_h^2 = \rho_f (v_f-v_h)^2$, which can be solved for $v_h$
\begin{equation}
\frac {v_f}{v_h}-1 \simeq
\left( \frac {4 \pi \delta r_s^2 v_f \rho_s}{\dot M_f} \right)^{1/2}.
\label{eq:vh1}
\end{equation}
The time the FFS jets crosse this shell and break out of it is given by (for $v_f \gg v_h$)
\begin{equation}
t_p \simeq \frac {\Delta r_s}{v_h} \simeq
\frac{\Delta r_s}{v_f}
\left( \frac {4 \delta \pi r_s^2 v_f \rho_s}{\dot M_f} \right)^{1/2}.
\label{eq:tp1}
\end{equation}
The mass residing in the surrounding of the disk is $M_s \simeq 4 \pi r_s^2 \Delta r_s \rho_s$.
Taking $\Delta r_s \simeq r_s$, equation (\ref{eq:tp1}) becomes
\begin{equation}
t_p \simeq
\delta^{1/2}
\left( \frac{r_s}{v_f} \right)^{1/2}
\left( \frac {M_s}{\dot M_f} \right)^{1/2}.
\label{eq:tp2}
\end{equation}

If there is no change in the system, the FFS jets will rapidly penetrate the surrounding
gas.
For example, let us consider a main sequence companion to an AGB star.
Scaling quantities accordingly and taking from now on ${\Delta r_s}=r_s$, and $r_s$ the
size of the orbital separation, equation (\ref{eq:tp2}) reads
\begin{equation}
t_p \simeq 3
\delta^{1/2}
\left( \frac{r_s}{1 \AU} \right)^{1/2}
\left( \frac{v_f}{500 \km \s^{-1}} \right)^{-1/2}
\left( \frac {M_s}{0.01 M_\odot}  \right)^{1/2}
\left( \frac {\dot M_f}{10^{-5} M_\odot \yr^{-1}} \right)^{-1/2} \yr.
\label{eq:tp3}
\end{equation}
For an accreting WD companion $v_f$ is $\sim 10$ times larger, but we can take
lower mass loss rate in the FFS jets, so the penetration time is the same.
Namely, in $\sim 1-10$~years the FFS jets break out.
This is a short time, and the FFS jets cannot accelerate much mass out of the system.
For scaling parameters used in equation (\ref{eq:tp3}) the velocity of the SMW jets
can be found from energy conservation, which gives $v_{\rm SMW} \simeq 27 \km \s^{-1}$.
This is too slow to be define as a jet in progenitors of PNs. The interaction time must
be longer in this case.
For an accreting WD $v_f \simeq 5000 \km \s^{-1}$, and for the same other parameters as in
equation (\ref{eq:tp3}) one finds $t_p \sim 1 \yr$, and $v_{\rm SMW} \simeq 150 \km \s^{-1}$.
I am seeking a more efficient process, which requires a longer interaction time.
Indeed, under the assumption of this study there is a constant change over a time scale $\tau_s$:
either mass is resupplied to the surrounding gas (from the donor and from the disk wind),
or the accreting body moves relative to the surrounding gas, e.g., an accreting star
orbit the mass donor star.
The condition for the FFS jets not to penetrate the surrounding gas is that the variation
time scale be shorter than the penetration time $\tau_s < \tau_p$.

Let the mass resupply rate to the surrounding gas be $\dot M_s \simeq M_s/ \tau_s$.
Using equation (\ref{eq:tp2}) the condition $\tau_s < \tau_p$ becomes
\begin{equation}
\tau_s \la
\delta
\frac{r_s}{v_f}
\frac {\dot M_s}{\dot M_f} =
\frac{\delta}{\eta}
\frac{r_s}{v_f}
\frac {\dot M_s}{\dot M_{\rm acc}},
\label{eq:taus}
\end{equation}
where in the second equality the ratio of the mass ejection rate in the FFS jets to the mass accretion rate,
$\eta= \dot M_f/ \dot M_{\rm acc}$, was defined.
We further define the velocity of the accreting body relative to the mass donor body, e.g.,
the mass donor star in binary systems or the intracluster gas in clusters of galaxies, $v_{\rm rel}$.
Then $\tau_s \simeq r_s/v_{\rm rel}$,and we can cast equation (\ref{eq:taus})
to be a condition on $\dot M_s$
\begin{equation}
\frac {\dot M_s}{\dot M_{\rm acc}} \ga
\frac {\eta}{\delta}
\frac {v_f}{v_{\rm rel}}.
\label{eq:ms1}
\end{equation}

\subsection{Studied cases}
I will substitute typical numbers in the expressions derived in section 2.2
for four cases.
In all case I will assume that $\sim 10 \%$ of the accreted mass is launched in
the FFS jets, i.e. $\eta \simeq 0.1$, and that the the FFS jets interact with the
surrounding gas before they are collimated, such that $\delta \sim 1$.
Some of the material in the FFS jets will interact with the dense disk wind, reducing the
demand on the mass in the surrounding medium $M_s$.
For that, in what follows I scale the equations with $\delta=1$.

\subsubsection{Planetary Nebulae}
The first two cases are of an accreting WD or a main sequence star orbiting close
to an AGB star $a \simeq 1-10 \AU$,
such as expected in some progenitors of bipolar PNs.
Typical relative orbital velocity of the two stars is in
the range $ v_{\rm Kep} \sim 30-10 \km \s^{-1}$, which is larger than the
AGB wind speed of $\sim 10 \km\s^{-1}$.
I therefore take  $v_{\rm rel}$ to be slightly larger than the average value
of $v_{\rm Kep}$.
For the FFS jets velocity I substitute the escape velocity.
Condition (\ref{eq:ms1}) reads
\begin{equation}
\left( \frac {\dot M_s}{\dot M_{\rm acc}} \right)_{\rm WD-AGB} \ga 20
\frac{1}{\delta}
\left( \frac {\eta}{0.1} \right)
\left( \frac {v_f}{5000 \km \s^{-1}} \right)
\left( \frac {v_{\rm rel}}{25 \km \s^{-1}} \right)^{-1}  ,
\label{eq:mswd}
\end{equation}
for an accreting WD, and
\begin{equation}
\left( \frac {\dot M_s}{\dot M_{\rm acc}} \right)_{\rm MS-AGB} \ga 2
\frac{1}{\delta}
\left( \frac {\eta}{0.1} \right)
\left( \frac {v_f}{500 \km \s^{-1}} \right)
\left( \frac  {v_{\rm rel}}{25 \km \s^{-1}} \right)^{-1} ,
\label{eq:msms}
\end{equation}
for an accreting main sequence star.

Equation (\ref{eq:msms}) shows that for the formation of SMW jets the mass donor AGB star should
transfer mass such that the mass slowly flowing in the accretion disk vicinity is about
equal to or larger than the mass accreted by the main sequence companion.
This will most likely to occur during a period of very high mass loss rate by the
AGB star, such that the accretion disk cannot `accept' all the mass lost by the AGB, and a steady state
is not reached.
Equation (\ref{eq:mswd}) show that a WD companion can form SMW jets when it accretes only few percent
of the mass lost by the AGB donor, with the condition that the rest of the mass lost by
the AGB donor slowly flows in the accretion disk vicinity;
here as well a period of high mass transfer rate is required.

\subsubsection{The Great Eruption of $\eta$ Carinae}
\label{sec:eta}
$\eta$ Carinae is a very massive binary stellar system that underwent a twenty years long eruption, termed the
Great Eruption, about 160 years ago (Davidson \& Humphreys 1997). A massive nebula of
$\sim 12 M_\odot$ (Smith et al. 2003) was expelled during the Great Eruption, now
forming the Homunculus, an expanding bipolar nebula around $\eta$ Car (Morse et al. 1998);
It is widely accepted now that $\eta$ Carinae is a massive binary system with an orbital
period of 5.54 yr, as first suggested by Damineli (1996).
The more massive component of the $\eta$ Car binary system will be referred to here as the primary,
while the more compact companion, probably an O-type star, will be referred to as the secondary.
There are two basic types of models for the formation of the Homunculus.
One based on axisymmetrical mass loss from the primary star (e.g., Smith 2006), and
the second, the binary model, is based on mass transfer from the primary
to the companion during the Great Eruption.
In the binary model (Soker, 2001, 2004b, 2005, 2007) an accretion disk was formed around
the companion, resulting in two jets that formed the Homunculus

Based on the results of Smith et al. (2003) and Smith (2006) of the velocity profile
and a mass of $\ga 12 M_\odot$,  I estimated the following
parameters of the Homunculus (Soker 2007).
The total energy in the Homunculus is
$E_h=3 \times 10^{49} \erg=1.5 \times 10^6 M_\odot \km^2 \s^{-2}$,
and the momentum discharge (adding the value of the momentum along all direction)  is
$5600 M_\odot \km \s^{-1}$.
In that paper (Soker 2007) I required that $4 M_\odot$ was launched by the secondary
at a velocity of $1000 \km \s^{-1}$, and a mass of $8 M_\odot$ was blown to the Homunculus
directly from the primary at a speed of $100 \km \s^{-1}$.
However, launching $4 M_\odot$ by the secondary requires  a large accretion rate by
the secondary, typically $20-40 M_\odot$, although I assumed there that only
$\sim 8 M_\odot$ were accreted.
Accretion rate that is only twice as high as the ejection rate, although possible,
is uncomfortable. This accreted mass is required to explain the total
energy budget of the Great Eruption according to the binary model (Soker 2007).

In the new model for the formation of SMW jets more comfortable parameters can be used
than those mentioned above.
The secondary accreted $\sim 8 M_\odot$ as is required to explain to total energy budget of
the Great Eruption (Soker 2007).
The accretion disk launched then the FFS jets at a speed comparable to its present wind
speed as deduced from X-ray observations (e.g., Corcoran et al.\ 2001;
Pittard \& Corcoran 2002; Akashi et al. 2006), $v_f \simeq 3000 \km \s^{-1}$.
While in Soker (2007) I assumed that most ($\sim 2/3$) of the mass in the Homunculus
came directly from the primary, here I assumed instead than most of it was
accelerated by the FFS jets, such that SMW jets were formed.
The average speed of the gas in the Homunculus is $v_{\rm hom} \sim 500 \km \s^{-1}$
(Smith 2006).
>From energy conservation the mass ejected in the FFS jets was
$\ga 12 (v_{\rm hom}/v_f)^2 M_\odot > 0.3 M_\odot$, as some energy is lost to radiation.
I take this mass to be $0.4 M_\odot$. Then, the accreted mass of $ 8 M_\odot$ can easily
supply this mass, i.e., $\eta \simeq 0.05$.

The relative orbital speed of the two stars in $\eta$ Car changes by more than an order
of magnitude along the highly eccentric orbit. The more relevant relative velocity is that
of the slow wind blown by the primary during the Great Eruption. In the binary model it is
assumed that this speed was much lower than the present primary wind speed
($\sim 500 \km \s^{-1}$), and it is taken to be $\sim 100 \km \s^{-1}$ (Soker 2005).

Using the parameters from the discussion above, condition (\ref{eq:ms1}) for the
Great Eruption in $\eta$ Car reads
\begin{equation}
\left( \frac {\dot M_s}{\dot M_{\rm acc}} \right)_{\rm GE} \ga 1.5
\frac{1}{\delta}
\left( \frac {\eta}{0.05} \right)
\left( \frac {v_f}{3000 \km \s^{-1}} \right)
\left( \frac {v_{\rm rel}}{100 \km \s^{-1}} \right)^{-1} .
\label{eq:mseta}
\end{equation}
In the binary model $M_s \sim 12 M_\odot$ and $M_{\rm acc} \simeq 8 M_\odot$.
These parameters are compatible with the model for the formation of the SMW jets
that formed the Homunculus.

\subsubsection{Cooling Flows in clusters of Galaxies}
\label{sec:cf}
There are two lines of arguments for SMW jets in cooling flow clusters.
Many of the X-ray deficient bubbles in cooling flow galaxies and clusters of galaxies
have more or less spherical structure, reside very close to the center of
the cluster (or galaxy), and are fully or partially surrounded by a dense shell;
Perseus (Fabian et al. 2000) and A~2052, (Blanton et al. 2001) being the best
examples.
These more or less spherical bubbles are termed `fat bubbles'.
In previous papers (e.g., Soker 2004a) I proposed that SMW jets can inflate fat bubbles.
This was shown to be the case with 2D gasdynamical numerical simulations
(Sternberg et al. 2007).
Typically, the half opening angle should be $\alpha \ga 50^\circ$, the jet speed
is $v_j \sim 3000-3\times 10^4 \km \s^{-1}$, and the mass loss rate in the two jets
$\dot M_{\rm SMW} \simeq 1-50  \dot M_\odot \yr^{-1}$.
Massive jets in cooling flow clusters were considered before but not with
wide opening angle (e.g., Begelman \& Celotti 2004; Binney 2004; Omma et al. 2004;
Heinz et al. 2006).

The second motivation comes from the required feedback heating in cooling flow
clusters (Soker \& Pizzolato 2005; Pizzolato \& Soker 2005).
Soker \& Pizzolato (2005) propose that a large fraction or even most of the gas that cools
to low temperatures in cooling flow clusters gains energy directly from the
central black hole, and is injected back to the ICM.
This is exactly the process in the proposed mechanism for the formation of SMW jets.

Other lines of arguments, mainly from observations, suggest than AGN
(not necessarily in cooling flow clusters) can have SMW jets (de Kool et al. 2001;
Crenshaw \& Kraemer 2007; Behar et al. 2003; Kaspi \& Behar 2006).

The velocity of the central cD galaxy relative to the cluster mean
is in the range $ \sim 0-400 \km \s^{-1}$ (e.g., Malumuth 1992).
The accreted matter can have higher velocity due to random motion in the ICM.
I take therefore $v_{\rm rel} \simeq 300 \km \s^{-1}$.
The FFS speed is scaled with the light speed $c$, such that $\eta$ stands for the
conversion factor of accreted energy to injected energy.
Condition (\ref{eq:ms1}) reads for cooling flow clusters of galaxies
\begin{equation}
\left( \frac {\dot M_s}{\dot M_{\rm acc}} \right)_{\rm cluser} \ga 100
\frac{1}{\delta}
\left( \frac {\eta}{0.1} \right)
\left( \frac {v_f}{c} \right)
\left( \frac {v_{\rm rel}}{300 \km \s^{-1}} \right)^{-1} .
\label{eq:mscluster}
\end{equation}

For a typical power of $\sim 5 \times10^{43}-5 \times 10^{45} \erg \s^{-1}$
in inflating bubbles (Rafferty et la. 2006) and $\eta=0.1$ the mass accretion
rate is $\sim 0.01-1 M_\odot\yr^{-1}$.
For the formation of SMW jets condition (\ref{eq:mscluster}) requires that the
mass ejection rate in the SMW be $\ga 1-100 M_\odot \yr^{-1}$.
This is compatible with the requirement for inflating fat bubbles (Sternberg et al. 2007)
and the demand for mass cycle in the moderate cooling flow model (Soker \& Pizzolato 2005).

\section{THE CONSTRAINTS ON THE FAST FIRST STAGE (FFS) JETS}
\label{sec:FFS}

The model for the formation of the MSW (massive-slow-wide) jets requires that the
FFS (fast-first-stage) jets be shocked, and that the postshock hot gas will accelerate
the mass $M_s$ in the surrounding medium.
This is schematically drawn in Figure 2.
For this mechanism to work the radiative cooling
time of the postshock FFS jets material should be longer than the acceleration time.

\subsection{Optically thin medium: PNs and AGNs}
As before, let the FFS jets have a total (to both sides of the accretion disk)
mass outflow rate of $\dot M_f$, a velocity $v_f$, and cover a solid angle of
$4 \pi \delta$ (on both sides of the disk together).
The FFS jets post shock density is
$\rho_b=4 \rho_f =  {\dot M_f}/{\pi \delta r_b^2 v_f}$, where $r_b$ is the
distance of the shock from the center.
The post shock temperature is $kT=(3/16)\mu m_H v_f^2$, with the symbols having their
usual meaning.
The radiative cooling time of an optically thin medium under a constant pressure is
\begin{equation}
t_{\rm rad} =\frac{5}{2}  \frac {n k T}{n_e n_p \Lambda} =
100  \delta %
\left( \frac{r_b} {1 \AU} \right)^2
\left( \frac{v_f} {5000 \km \s^{-1}} \right)^3
\left( \frac{\dot M_f} {10^{-8} M_\odot \yr^{-1}} \right)^{-1}
\left( \frac{\Lambda} { 10^{-22} \erg  \cm^3 \s^{-1}} \right)^{-1}
\yr
\label{eq:tco}
\end{equation}
where $n$, $n_e$ and $n_p$ are the total, electron and proton numbers density respectively,
and $\Lambda$ is the cooling function.
The mass loss rate of the FFS jets and speed were chosen to meet FFS launched by a WD.
The SMW jets formed by such FFS jets can have $v_{\rm SMW} \simeq 500 \km \s^{-1}$ and
$\dot M_{\rm SMW} \simeq 10^{-6} M_\odot \yr^{-1}$, and can lead to the formation of
some bipolar PNs (Akashi 2007; Akashi \& Soker 2008).
As in the previous section I substitute $\delta=1$.
The flow time is the region size divided by the MSW jets speed, rather than the FFS jets speed
\begin{equation}
t_{\rm flow}=\frac {r_b}{v_{\rm SMW}}= 0.01
\left( \frac{r_b} {1 \AU} \right)
\left( \frac{v_{\rm SMW}} {500 \km \s^{-1}} \right)^{-1} \yr.
\label{eq:tfl}
\end{equation}
I assume that all the kinetic energy of the FFS jets is transfer to the SMW jets, such that
$\dot M_{\rm SMW} v_{\rm SMW}^2 \simeq \dot M_f v_f^2$. Taking the SMW jets mass to
be equal to the surrounding gas mass $\dot M_{\rm SMW} \simeq \dot M_s$, and
using the definition of $\eta$, we find
$v_{\rm SMW} \simeq v_f (\eta \dot M_{\rm acc}/{\dot M_s})^{1/2} \sim 0.1 v_f$.

Comparing equations (\ref{eq:tco}) and (\ref{eq:tfl}) it is clear that the FFS blown by WD
companions to AGB stars as required in the binary model for shaping bipolar PNs
have long radiative cooling time, as required.
However, for FFS blown by main sequence stars, for which $v_f \sim 500 \km \s^{-1}$, the radiative
cooling time is not much longer than the outflow time.
We can still get SMW jets for ${\dot M_f}\simeq {10^{-8} M_\odot \yr^{-1}}$,
$v_f \sim 500 \km \s^{-1}$, and $v_{\rm SMW} \sim 100 \km \s^{-1}$, but the parameters space
volume is small.
The conclusions is that in PNs energetic SMW jets are most likely formed by accreting WD companions to
the AGB progenitors. Main sequence accreting companions can also blow SMW jets, but either they will be
less energetic, or they will be the FFS jets themselves; in that case they will not be slower than the
escape speed from the accreting main sequence star.

For jets from super-massive black holes for which the Schwarzschild radius is $\sim 10 \AU$,
I take $r_b \sim 10^4 \AU \simeq 0.05 \pc$.
According to section \ref{sec:cf} for cooling flow clusters
$\dot M_f \sim 0.001-0.1 M_\odot\yr^{-1}$. Using an appropriate value for the cooling
faction and a speed close to the speed of light, $\sim 10^8 \cm \s^{-1}$,
I find the radiative cooling time to be $t_{\rm rad} \sim 10^9-10^7 \yr$.
The flow time is only $t_{\rm flow} \sim 10 \yr$.
The conclusion is that in cooling flow clusters the radiative cooling time is much
longer than the flow time 
and the FFS jets can form SMW jets,
if condition (\ref{eq:mscluster}) is met as well.

\subsection{Optically thick medium: The Great Eruption of $\eta$ Car}
Following section \ref{sec:eta} $\sim 0.4 M_\odot$ were blown in the FFS jets during the
$\sim 20 \yr$ Great Eruption, namely, $\dot M_f  \simeq 0.02 M_\odot \yr^{-1}$.
The velocity is $v_f \simeq 3000 \km \s^{-1}$.
The radiative cooling time would be much shorter than the flow time according to equations
(\ref{eq:tco}) and (\ref{eq:tfl}). However, for such a flow the medium is not optically thin any more.
The optical depth of the shocked FFS jets is
\begin{equation}
\tau_f \simeq \kappa \rho_b r_b \simeq 30
\frac{1}{\delta}
\left( \frac{r_b} {1 \AU} \right)^{-1}
\left( \frac{\dot M_f} {0.02 M_\odot \yr^{-1}} \right)
\left( \frac{v_f} {3000 \km \s^{-1}} \right)^{-1},
\label{eq:tauf}
\end{equation}
where the opacity $\kappa$ is that of scattering on free electrons.
The optical depth of the surrounding $M_s$ mass is much larger.
The velocity is only $\sim 100 \km \s^{-1}$, and the mass loss rate
is $\times 30$ times larger ($\sim 12 M_\odot$ lost in $\sim 20 \yr$).
At apastron the relevant distance can be larger, though.
Over all, the optical depth of the surrounding gas in conditions appropriate
to the Great Eruption o f$\eta$ Car according to the model is $\tau \sim 10^3$.

The relevant cooling time for $\tau \gg 1$ is the photon diffusion time over
a distance r: $t_{\rm cool} \simeq \tau r/c$.
This should be compared to the outflow time $t_{\rm flow} = r/v$.
For $\eta$ Car the SMW jets have $ v \sim 500 \km \s^{-1}$,
and hence $t_{\rm cool}/t_{\rm flow} \simeq \tau v/c \sim 1$.
The FFS jets' shocked gas has time to accelerate the surrounding gas and form the SMW jets.
However, radiative losses are not negligible, and not all of the FFS energy will
be transfer to the SMW jets. This is still enough to account for the expanding Homunculus
(Soker 2007).

The total radiated energy from the Great Eruption is $E_{\rm GE-rad} = 3\times 10^{49} \erg$
(Humphreys et al. 1999), which is equal to the kinetic power in the Homunculus $E_h$.
Indeed the entire energy radiated by the Great Eruption, whether from the shocked
FFS jets or directly from the two stars and the accretion disk, exerted radiation
pressure on the surrounding gas.
The photosphere of the erupting system was at a distance of $\sim 20 \AU$
(Davidson \& Humphreys 1997), showing indeed that radiation was
efficient in accelerating the gas.
Radiation only from the two stars without the accretion disk and the FFS jets
would have lead to a more spherical ejection, rather than a bipolar nebula (the Homunculus).
The FFS jets are required for the formation of a bipolar structure according to the
model presented here.

\section{SUMMARY}
\label{sec:summary}

Motivated by supporting arguments for the existence of slow-massive-wide (SMW) jets
in planetary nebulae (PNs), during the Great Eruption of $\eta$ Car, and in some AGNs in galaxies
and in cooling flow clusters (see section 1), I proposed a model for their formation.
By 'slow' I refer to jets much slower than the escape velocity from the accreting object,
by `massive' I refer to jets with their mass outflow rate of the same order of magnitude or larger
than the accretion rate onto the accreting object, and by `wide' I refer to
a half opening angle of $\alpha \ga 20 ^\circ$.

In this model the energy still comes from accretion onto the compact object.
Jets with velocity of the order of the escape velocity from the accreting object
and with mass outflow rate of $\sim 1-20 \%$ of the accretion rate are launched by
the inner zone of the accretion disk as in most popular models for jet launching.
In the present model these are termed fast-first-stage (FFS) jets.

The first condition for the formation of SMW jets is that a massive surrounding gas will
block the free expansion of the FFS jets; this was studied in section 2.
If the FFS jets penetrate this surrounding mass $M_s$,
they will be collimated by it, and narrow jets will be formed. If, on the other hand, the
FFS jets cannot penetrate the surrounding mass, they will accelerate it.
A similar type of flow was simulated by Sternberg et al. (2007) and
Sutherland \& Bicknell (2007), although with very different parameters.
Although different quantitatively and qualitatively, these simulations show that
fast jets can form a wide slower outflow.

The surrounding mass along the propagation direction of the FFS jets must be replenish
on a time scale $\tau_s$ shorter than the penetration time $\tau_p$.
In the present paper this surrounding mass is coming from the same source as that of the
accretion disk.
When the mass transfer rate to the accreting object is very high and violent, not all
of the mass will reach the accretion disk, and the mass that reaches the disk cannot
settled into a steady state flow.
Some of the mass ends up in regions above and below the disk.
This can occur, for example, when in a binary system the mass donor star is in eruption,
or in a phase of a high mass cooling rate in cooling flow clusters.

The first condition can be expressed as constraint on the ratio of mass flowing to the
surrounding of the accretion disk and to the central object.
I derived this condition for progenitors of PNs when the mass donor
star is an AGB star, and the accreting companion is a WD (eq. \ref{eq:mswd}) and
and a main sequence star (eq. \ref{eq:msms}), for the formation of the Homunculus
(the bipolar nebula) of $\eta$ Car during the great eruption (eq. \ref{eq:mseta}),
and for cooling flow clusters (eq. \ref{eq:mscluster}).
As evident from these equations,  most of the mass involved in the process
resides in the surrounding of the accretion disk and is eventually expelled from the
system.
As stated above this requires a non-steady state mass transfer process.

As I discussed in section 2.3, massive jets better fits the properties of the Homunculus
in $\eta$ Car, and can account for the inflation of fat (more or less spherical) bubbles
in cooling flow clusters.

The second condition for the formation of SMW jets is that the radiative energy losses
be small; this was studied in section 3.
This implies that the radiative cooling time scale of the shocked FFS jets gas
be longer than the acceleration (flow) time scale.
For PN progenitors with WD accreting companions and in cooling flow clusters this condition
is met by a large margin because the radiative cooling time of optically thin medium in
these systems is much longer than the flow time scale.
In the case of PN progenitors with main sequence accreting stars, this condition is
marginal. In that case it is possible that the MS star itself will directly launched wide
jets (although they cannot be massive).
In the case of the Great Eruption in $\eta$ Car the medium was optically thick, and the
photon diffusion time was about equal to the flow time.
This means that any radiation, whether from the shocked FFS jets material or
from the accreting object, will accelerate the SMW jets. However, the FFS jets are
required to form the bipolar outflow.

More detail studies of the formation mechanism of SMW jets require numerical
gasdynamical simulations.
The parameter space of the flow structure studied here is very large.
However, the basic flow structure and the question of whether the mechanism proposed here
for the formation of SMW jets can work, can be explored with existing numerical codes.
I hope that this preliminary study will motivate deeper search into the formation
mechanism(s) of SMW jets.

\acknowledgements
This research was supported by the Asher Space Research Institute in the Technion.

\end{document}